\magnification=1200

\def\.{\mathaccent 95}

\def\ga{\gamma}

\def\ep{\epsilon}

\def\rh{\rho}

\def\om{\omega}

\def\Om{\Omega}

\def\frac#1#2{{\textstyle{{#1}\over {#2}}}}

\def\lsim{\mathrel{\rlap{\lower4pt\hbox{\hskip1pt$\sim$}}
    \raise1pt\hbox{$<$}}}
\def\gsim{\mathrel{\rlap{\lower4pt\hbox{\hskip1pt$\sim$}}
    \raise1pt\hbox{$>$}}}
\def\sqr#1#2{{\vcenter{\vbox{\hrule height.#2pt
         \hbox{\vrule width.#2pt height#1pt \kern#1pt
         \vrule width.#2pt}
         \hrule height.#2pt}}}}

\newbox\grsign \setbox\grsign=\hbox{$>$} \newdimen\grdimen \grdimen=\ht\grsign
\newbox\simlessbox \newbox\simgreatbox
\setbox\simgreatbox=\hbox{\raise.5ex\hbox{$>$}\llap
     {\lower.5ex\hbox{$\sim$}}}\ht1=\grdimen\dp1=0pt
\setbox\simlessbox=\hbox{\raise.5ex\hbox{$<$}\llap
     {\lower.5ex\hbox{$\sim$}}}\ht2=\grdimen\dp2=0pt

%
%

\def\ref#1  {\noindent \hangindent=24.0pt \hangafter=1 {#1} \par}
\def\doublespace {\smallskipamount=6pt plus2pt minus2pt
                  \medskipamount=12pt plus4pt minus4pt
                  \bigskipamount=24pt plus8pt minus8pt
                  \normalbaselineskip=24pt plus0pt minus0pt
                  \normallineskip=2pt
                  \normallineskiplimit=0pt
                  \jot=6pt
                  {\def\smallskip {\vskip\smallskipamount}}
                  {\def\medskip   {\vskip\medskipamount}}
                  {\def\bigskip   {\vskip\bigskipamount}}
                  {\setbox\strutbox=\hbox{\vrule 
                    height17.0pt depth7.0pt width 0pt}}
                  \parskip 12.0pt
                  \normalbaselines}
\def\tm{\times}
\def\ra{(Meegan et al., 1994) }
\def\rb{(Koveliotou, 1994) }
\def\rc{(Nemiroff et al., 1994) }
\def\rd{(van den Heuvel, 1986)}
\def\re{(Usov, 1992) }
\def\rf{(Usov, 1994) }

\def\rh{(Blandford, 1993) }
\def\ri{(Ostriker \& Gunn, 1969) }
\def\rj{(Friedman, 1983) }
\def\rk{(Thorne et al., 1986) }
\def\rl{(Vermeulen, 1993) }
\def\rrm{(e.g. Krolik \& Pier, 1991) }
\def\rn{(Goldreich \& Julian, 1969) }

\def\rp{(Asseo et al., 1978) }
\def\rq{(Lamb et al., 1983) }
\def\rr{(e.g. Narayan \& Popham, 1989) }

\magnification=1200
\centerline{\bf  RELATIVISTIC PRECESSING JETS}
\centerline{\bf AND COSMOLOGICAL GAMMA-RAY BURSTS}
\medskip
\centerline{\bf Eric G. Blackman}
\smallskip
\centerline{Institute of Astronomy, Madingley Road, Cambridge, CB3 OHA,
England}
\smallskip
\centerline{\bf Insu Yi}
\smallskip
\centerline{Institute for Advanced Study, Olden Lane, Princeton, NJ, 08540}
\smallskip
\centerline{and}
\smallskip
\centerline{\bf George B. Field}
\smallskip
\centerline {Harvard-Smithsonian Center for Astrophysics, 60 Garden St., Cambridge, MA, 02138}

\doublespace
\medskip
\centerline{(accepted to {\it Ap.J. Lett.})}

$$\bf ABSTRACT$$
 
We discuss the possibility that gamma-ray bursts may result from
cosmological relativistic blob emitting neutron star 
jets that precess past the line of sight.
Beaming reduces the energy requirements, so that the jet emission can 
last longer than the observed burst duration. 
One precession mode maintains a short duration time scale, while a
second keeps the beam from returning to the line of sight, consistent with
the paucity of repeaters.
The long life of these objects reduces the number required for production
as compared to short lived jets.  Blobs can account for the
time structure of the bursts.  Here we focus largely on kinematic
and time scale considerations of beaming, precession, and blobs--issues
which are reasonably independent of the acceleration and jet collimation
mechanisms.  We do suggest that large amplitude electro-magnetic waves 
could be a source of blob acceleration.

$${\bf Subject\ headings}:{\rm\ gamma-rays:\ bursts;\ ISM:\ jets\ \& \ outflows;\ pulsars: general}$$

\vfill
\eject
$\bf I.\ Introduction$

Gamma-ray bursts (GRB) are an isotropically distributed class of
transient gamma-ray emitters \ra showing fluxes $\ge\sim 10^{-7}{\rm erg\
sec^{-1}cm^{-2}}$.  All except three have not been seen to repeat and thus
have a repetition time scale  $\tau_{rep} > 20\ {\rm yr}$.  GRB have
durations $0.1\ {\rm sec}\le \tau_{dur} \le 1000\ {\rm sec}$.
The time profiles vary substantially but statistically show 
rise times shorter than fall times \rc.  The repeaters
show smoother profiles, but all show variabilities  
on the smallest $<\sim 1\ {\rm ms}$ observable scales.  GRB spectra are 
non-thermal \ra.

Here we introduce the idea that GRB may result from 
cosmological blob-emitting pulsars formed 
(PSRs) from accretion induced collapse \rd $\ $ of white dwarfs (WDs).  
Two types of emitting PSRs can result from the collapse process: those that
spin faster than their gravitationally unstable limit which we call
super-critical rotators (SPR) and those that spin
below this limit which we call sub-critical rotators (SBR).
The two objects are distinct in that the former loses most of its
rotational energy through gravitational radiation during a very short time.
We investigate the possibility 
that these PSRs are a source of GRB when
their emission accelerates blobs relativistically within a collimated
jet.  Two precession (PN) modes associated with the binary companion then
conspire to move the jet out of the line of sight, the first
accounting for a short $\tau_{dur}$, and the second
keeping $\tau_{rep}> 100\ {\rm yr}$ (Blackman, 1995).  

There have been previous associations of pulsars with GRBs
(e.g. Usov, 1992; Usov, 1994; Fatuzzo \& Melia, 1993; Thompson, 1994).
In our scheme  however, beaming reduces the energy requirements, and so we do not require the anomalously strong magnetic fields of $\sim 10^{15}{\rm G}$ 
(e.g. Usov, 1992; Usov, 1994, Thompson, 1994).  Beaming also allows 
the objects to emit for a period much longer than their observed durations.
We will point out that this reduces the number of bursts
required to be produced per year when compared to a model for which the
anisotropic emitter has a lifetime $=\tau_{dur}$.  

In sections II-IV we discuss  the role of
beaming, precession, and blobs for GRB.  
These ingredients are independent of the particular 
emission and jet collimation mechanisms, and thus the relevant discussion
is largely  kinematic.  We  suggest a possible emission
mechanism in section V, but other acceleration processes could also
be subject to beaming and precession.  
Using the particular mechanism of section V, we
discuss how the observed kinematic requirements constrain the
PSR properties.  We find that typically, $\tau_{life}\sim 10^{4} {\rm yr}$
with $B_{s0} \sim  10^{12}{\rm G}$ and $\Om_s \sim 10^4 {\rm
sec^{-1}}$, 
where $\tau_{life}$ is the gamma source lifetime, $B_{s0}$ is the surface 
magnetic field, and $\Om_s$ is the PSR
angular velocity.
In section VI we discuss the required rate of production and 
compare our GRB sources to SS433.

$\bf II.\ Kinematics\ and\ Pulsar\ Classes$

If  GRB  emission comes from 
collimated blobs moving along the line of sight, 
the blob power source  must have an intrinsic luminosity of \ra
$$L_{int}\sim \gamma^{-2}L_{ob s}\sim \ga^{-2} 10^{51}{\rm erg/sec},\eqno(1)$$
where $L_{obs}$ is the observed luminosity received per solid angle.  

If the blob energy source is PSR rotational kinetic
energy, $L_{int}$ is given by the magnetic dipole formula \re
$$L_{int}\sim f L_{dip}\sim [10^{44}{\rm erg/sec}]~f~(\Om_s/10^4{\rm
sec^{-1}})^4(B_{s0}/10^{12}G)^2 (R_s/10^6{\rm cm})^6\ ,\eqno(2)$$
where $R_s$ is the PSR radius, and $f<1$ is the fraction of luminosity
going into the observed accelerated particle emission.
The gravitational radiation luminosity for a PSR is given by \ri
$$L_{gr}\sim [10^{33}{\rm erg/sec}](\ep/10^{-11})^2 
(E_{kin}/10^{53}{\rm erg})^2(\Om_s/10^4{\rm sec^{-1}})^2,\eqno(3)$$
where $E_{kin}$ is the rotational energy, and
$\ep$ is the PSR ellipticity \re with 
$\ep\sim 10^{-11}(B_{s0}/10^{12}G)^2,$ for  $\Om_s<\Om_{uns}$
and $\ep \sim10^{-1}$  for $\Om_s\ge\Om_{uns}$,
where $\Om_{uns}>\sim {\rm O}(10^4)$ sec$^{-1}$ is the  
angular velocity above which the PSR is gravitationally unstable \rj.  Thus 
the emission lifetime, $\tau_{life}$, satisfies
$$\tau_{life}= E_{kin}/(L_{int}+L_{gr}).\eqno(4)$$
When initially, $L_{gr}>\sim L_{int}$, we have a SPR whereas when
$L_{int}\ll L_{gr}$, we have a SBR.

The GRB PSRs would 
form from accretion induced collapse of a WD in a close
binary \rd.  
For an angular momentum conserving collapse, 
$$\Om_s
\sim 10^6(R_{wd}/10^9{\rm cm})^2(R_{s}/10^6{\rm cm})^{-2}\Om_{wd},\eqno(5)$$ 
where $\Om_{wd}$ and $R_{wd}$ are the WD angular velocity and radius.
For the SPR, $\Om_s>\Om_{uns}$, and the jump in $\ep$
at $\Om_s\sim \Om_{uns}$ means that there are two phases to the emission as
determined by $L_{gr}$.  We have from $(3)$ and $(4)$,   
$$\tau_{_{SPR}}\sim [0.1\ {\rm sec}](E_{kin}/10^{53}{\rm erg})^{-1}
(\Om_s/10^4{\rm sec^{-1}})^{-2}.\eqno(6)$$
For typical parameters, in a time $\sim\tau_{SPR}\lsim$ few sec, 
the ellipticity $\epsilon$ of the pulsar makes a transition and the 
electromagnetic $L_{int}$ decreases  with the rapid depletion of the
available dipole power (Yi \& Blackman, 1996). This results from a
 transition of a non-axisymmetric Jacobi ellipsoid to a
axisymmetric Maclaurin spheroid. For a uniform ellipsoid, the angular 
velocity actually increases during the transition (Chandrasekhar, 1987).  
Although the pulsar energy is depleted on a time scale $\sim\tau_{_{SPR}}$,
the exact time variation of the electromagnetic power during this phase
depends on several outstanding uncertainties such as the nature of the 
magnetic field (Usov 1992, Thompson 1994) and the pulsar's dynamical structure 
including the rotational velocity law inside the star (Yi and Blackman 1996). 
If the SPR jet loses its gamma-ray luminosity on $\tau_{_{SPR}}$, then
$\tau_{dur}\sim \tau_{SPR}\sim \tau_{life}$.
For SBR, $\Om_s<\Om_{uns}$ and $\ep$ is small.  
Here the PSR rotates stably, 
and from $(2)$ and $(3)$ we then have
$$\tau_{life}\sim E_{kin}/L_{int}\sim 10^9(E_{kin}/10^{53}{\rm erg})
(\ga/3\times 10^4)^2(L_{obs}/10^{51}{\rm erg/sec})^{-1}{\rm sec}.\eqno(7)$$ 
The jet formation time scale is difficult to estimate
but a characteristic dynamical time scale is that of the neutron
star rotation, which would be of order $10^{-6}$ seconds, which is the smallest
time scale in our study.

We emphasize that $\tau_{life}\gg\tau_{dur}$ is not a problem, 
as this is where
PN enters the picture.  PSR formation by accretion induced collapse
does not produce significant mass ejection \rd, and would leave the
binary companion as a PN agent.  The bursts  are  
seen only when the gamma-ray emitting cross section of their
jet precesses by the line of sight and we now discuss this further.

$\bf III.\ Role\ of\ Precession\ for\ SBR$

The sweep time scale for gamma emission $\tau_{swp}\sim\tau_{dur}$ depends on 
$\ga$ as well as the jet PN time scale; ultra-relativistic jet
blob emission is 
beamed in its outflow direction to an angle 
$\theta\sim \gamma^{-1}$.  This links $\tau_{dur}$ to the 
shortest PN time scale, $\tau_{pn1}$, since
$$1/2\pi\gamma\sim\tau_{dur}/\tau_{pn1}.\eqno(8)$$

The more cylindrical the jet channel in
which the gamma-ray emitting blobs move, the less stringent the
requirement on the beaming of the whole jet:  
For a conical jet, the entire jet must be beamed to an angle $\sim
\ga^{-1}$ which is $\sim 10^{-4}$.  
However, if the jet is cylindrical, then there is no such requirement
on the physical jet flow, only that the blobs produced have 
diameter of order the jet width.  
These arguments apply to the blobs that emit gamma-rays; 
a cylindrical gamma-ray emitting jet 
could be contained within a large conical jet that emits 
blobs of much smaller $\ga$'s.  Cylindrical jets have been studied
in the context of AGN (Heyvaerts \& Norman, 1989; Chiueh et al., 1992)

Although $\tau_{dur}$ results from the shortest PN time scale, 
$\tau_{rep}$ is determined by a
combination of PN time scales.  A secondary PN
can move the PN cone of $\tau_{pn1}$
so that the blob beam will not return to the line of sight on a
time scale $\tau_{pn2}>>\tau_{pn1}$.  For  
multiple PN modes to operate in this way, the
time scales must satisfy 
$\tau_{pn1}/\tau_{pn2}\ge\tau_{swp}/\tau_{pn1}\sim 1/2\pi\ga.$
For large $\ga$, a wide range of
PN time scales can thus account for $\tau_{rep}>>\tau_{dur}$.  

In our PSR-binary system, 
three important PN frequencies are associated with the binary
interaction \rk.  In increasing order of magnitude we denote these by
${\bf \Om}_T$, ${\bf \Om}_J$, and ${\bf \Om}_G$.  
The first comes from the Newtonian tidal torque.  
The second results from interaction between the PSR spin and 
the gravitomagnetic field associated with the companion spin. 
The third results from 1) PSR spin interaction
with the gravitomagnetic field associated with the companion orbit
and the external gravitational field of the companion, 2) space-curvature 
PN, and 3) a ``spin-orbit'' PN from an additional 
gravitomagnetic field induced by the orbital motion of the 
NS in the gravitational field of the companion.  
Note that gravitomagnetic PN of compact objects is 
analogous to PN of the magnetic moment of a
particle around its spin axis in the presence of a magnetic field \rk.  
The PN of the spin axis relative to the fixed stars is then given by
$$d{\bf J}_s/dt= ({\bf \Om}_{T}+{\bf \Om}_{J}+{\bf \Om}_{G})
\tm {\bf J}_s,\eqno (9)$$
where ${\bf J}_s$ is the PSR angular momentum, and $t$ is the time.
For a nearly maximally rotating PSR, the main contributions to the
three frequencies are then given by \rk 
$$\Om_{G}\sim [6\tm 10^{-8}{\rm sec^{-1}}]\left[3M_c+M_cM_s/(M_c+M_s)
\over 1.2M_\odot\right]\left(\Om_K\over 5\tm 10^{-3}{\rm sec^{-1}}\right)
\left(R_B\over 2\tm 10^{10}{\rm cm}\right)^{-1}\eqno(10)$$
$$\Om_J=0.5\Om_G(R_c/10^{10} {\rm cm})^{1/2}(R_B/2\tm 10^{10}{\rm cm})^{-1/2}\eqno(11)$$
$$\Om_{T}\sim 10^{-3}\Om_G(M_s/M_\odot)^{1/2}(R_B/2\times10^{10}{\rm cm})^{-1/2},\eqno(12)$$
where $M_s\sim M_\odot$ and $M_c\sim 0.5 M_\odot$ are characteristic 
masses of the PSR and companion, and $R_B$ and
$R_c$ are the binary and companion radii.
For the typical values shown,
$\tau_{dur}\sim \tau_{pn1}/2\pi\ga=\Om_G^{-1}/\ga\sim 100
{\rm sec}(\ga/10^5)^{-1}$ and
$\tau_{rep}= \tau_{pn2}=\pi \Om_T^{-1}\sim 2\tm 10^{11}{\rm sec}$, so that 
$\tau_{dur}<<\tau_{rep}$.

$\bf IV.\ Role\ of\ Blobs$

Although GRB statistically show asymmetric time profiles, these 
are not observed for all GRB \rc, and it is
thus best if a unifying model allows for a myriad of profiles.
Toward this end, we note that blob emission is a standard mode by
which relativistic motion is often observed in jets \rh, and in the present
context can allow a sweeping jet to produce asymmetric,
symmetric and multi-peaked profiles.
The observed GRB time profiles are then characterized by the
frequency of blob production $\om_{prod}$, and the blob decay time
$\tau_{dec}$, relative to the time scale for the jet channel to sweep
across our line of sight, $\tau_{swp}$. 
An asymmetric profile would result if $\om_{prod}^{-1}>\tau_{swp}$ and
$\tau_{dec}<\tau_{swp}$.  In this case, the rise and fall times would be
determined by the rise and decay time of the emitting  blob, as
the jet would be essentially stationary relative to the blob decay time.
A symmetric profile would result if either
$\tau_{dec}>\tau_{swp}$ or if $\om_{prod}$ is sufficiently 
large that the emission is nearly uniform in a sweep time.  In this
case, the rise and fall times are simply determined by $\tau_{swp}$.  
The presence of profiles of all types requires that roughly  
$$\tau_{dur}\sim\tau_{dec}\sim \tau_{swp} \sim \om_{prod}^{-1},\eqno(13)$$ 
even if the asymmetric profiles statistically dominate.
(In a specific blob production model, the environment of the
burst might create multiple blobs in a jet diameter broader than that
determined by the solid angle derived from the beaming of any
individual blob.  This would give multipeaked emission.)  
Condition $(13)$ is actually consistent with observations of SS433 \rl
which we further discuss later.   

$\bf V.\ Possible\ Acceleration\ Mechanism\ for\ Gamma-Ray\ Emission$

Avoiding runaway pair production that would make a blob optically
thick to gamma rays, requires $\ga>100$ \rrm.
Pair plasma blobs moving along the magnetic dipole axis 
with $\ga>\sim 10^4$ might be produced 
in PSR magnetospheres \rf by
large amplitude electro-magnetic waves (LAEM). 
Here, a fraction $f$, used in eqn. (2), of the LAEM wave
energy is absorbed by each blob which is then accelerated to
large $\ga$ and subsequently emits  radiation. 
We consider here a scenario for which
the blobs are observed only along a nearly cylindrical channel,
though we do not have a model for formation of this channel. 
We consider the blobs to be accelerated
by the fraction of the nearly spherical wave contained within
the channel cross section.  
Note that the channel is simply that for the gamma-ray blobs, there
could be lower energy emission at radii outside that of the gamma ray
blob channel axis.
(An alternative scheme for the SBR 
is one in which the LAEM are themselves focused entirely
into a narrow beam.)
The relevant beaming is simply the beaming of blob emission, as the blobs
are accelerated to large Lorentz factors.  

The (LAEM) propagate outside a radius, $r_{ff}$, where the density
drops below that which can sustain a Goldreich-Julian \rn charge
density, and thus where flux-freezing and force-free conditions are
broken.  This gives \rf
$$r_{ff}\sim 10^{12}{\rm cm}(B_{s0}/10^{12} G)^{1/2}
(\Om_s/10^4{\rm sec^{-1}})^{1/2},\eqno(14)$$
and the associated pair plasma number density 
$$n_{ff}\sim 10^{6}{\rm cm^{-3}}(R_s/10^6{\rm cm})(B_{s0}/10^{12}{\rm G})^{1/2}(\Om_s/10^4{\rm sec^{-1}})^{5/2}.\eqno(15)$$
Electron acceleration at  $r_{ff}$ is characterized by
the parameter $\sigma_{ff}$, defined by (Usov, 1994; Michel, 1984) 
$$\sigma_{ff}\equiv L_{int}/(mc^2{\dot N}_{ff})\sim 5 \tm
10^7(R_s/10^6{\rm cm})^5
(B_{s0}/10^{12}G)^{1/2}(\Om_s/10^4{\rm sec^{-1}})^{1/2},\eqno(16)$$
where ${\dot N}_{ff}=4\pi r_{ff}^2cn_{ff}$ is the electron flux 
and $m_e$ is the electron mass.
By solving the equations of motion for a particle in a pulsar wind zone
subject to electromagnetic forces,  it has been shown that 
relativistic electromagnetic 
waves of frequency $\Om_s$ can accelerate pair plasma to
(Michel, 1984, Asseo et al., 1978)
$$\ga \sim \sigma_{ff}^{2/3}\sim 10^5 (R_s/10^6{\rm cm})^{10/3}
(B_{s0}/10^{12}G)^{1/3}(\Om_s/10^4{\rm sec^{-1}})^{1/3},\eqno(17)$$
and the resulting emission is beamed within solid angle $\sim
\ga^{-2}$ from the direction of wave propagation \rp.
The characteristic emitted frequency of the synchro-Compton radiation 
is proportional to  $\ga^3$ or $\sigma_{ff}^2$ from (17) (like
curvature radiation, eg. Beskin et al., 1993).  In particular,  \rp
$$\om_{sc} \sim 10^{20}{\rm sec^{-1}}(\Om_s/10^4{\rm
sec^{-1}})(\sigma_{ff}/10^8)^2,\eqno(18)$$
with a tail to
$$\sigma_{ff}^{4/3}[eB_s(r_{ff})/(m_ec)]\sim 10^{24}{\rm sec^{-1}}
(B_{s0}/10^{12}{\rm G})^{7/6}(R_s/10^6{\rm cm})^{23/3}
(\Om_s/10^4{\rm sec^{-1}})^{1/6}.\eqno(19)$$

For GRB,  $\ga$ is important in determining  $\omega_c$ of (18)
and also in providing a fundamental 
reduction in the source energy requirements.  This reduces the required
magntidude of the magnetic field through $(1)$.
Specifically, given the observed $\tau_{dur}$,
$L_{obs}$ and $\om_{sc}$, the quantities 
$B_{s0}$, $\Om_s$ and $R_s$ can be estimated for a given $f$ and
for $\tau_{dur}\sim\tau_{swp}$.
Spin-orbit synchronization gives
$\Om_s=10^6\Om_K$ from $(5)$ so the PN frequency $(10)$ depends
on $\Om_s$. 
We take $M_s=5M_c=M_\odot$, $f\sim 0.01$,
$L_{obs}\sim 10^{51} {\rm erg/sec}$, $\om_{sc}\sim 10^{20}{\rm
{sec^{-1}}}$ and $\tau_{dur}\sim 100{\rm sec}$.
Then  $(1)$, $(8)$, and $(18)$, using $(10)$ are fit well by
$B_{s0}\sim 10^{12}G$, $\Om_s \sim \tm 10^{4}{\rm sec}$, and
$R_s\sim 0.9 \tm 10^6 {\rm cm}$.  In this case, from $(17)$ 
$\ga\sim  10^5$, and $\tau_{life}\sim 10^3 {\rm yr}$ from $(7)$.
  
If the SPR dim on a time scale $\tau_{SPR}$, PN is irrelevant.  
The equations to consider
are then $(1)$, $(18)$ and $(6)$ since for these particular SPR
$\tau_{dur}=\tau_{life}$.   For $f\sim 0.01$,  
$L_{obs}\sim 10^{51} {\rm erg/sec}$,
$\om_{sc}\sim 10^{20}{\rm sec}^{-1}$ and $\tau_{dur}\sim 0.05 {\rm
sec}$ the equations are well fit by
$B_{s0} \sim  10^{11} {\rm G}$, $\Om_s \sim 1.3\tm 10^{4}{\rm sec}$, and
$R_s\sim 1.3\tm 10^6 {\rm cm}$ for the unstable PSR.  For such SPR, 
$\ga\sim 10^5$.  
The smaller $B_{s0}$ and larger $\Om_s$ for SPR
compared to SBR are consistent with accretion induced collapse:
Only when the initial WD magnetic field $>\sim 10^6 G$, 
$\Om_{wd}$ can synchronize with the binary orbital frequency
$\Om_K$, within a Hubble time \rq. 
A reduced magnetic torque correlates a smaller field with a 
larger $\Om_{wd}$ and $\Om_s$ \rr.    
Whether the differences between SPR and SBR can be related
to the observed bimodal distribution in GRB durations \rb, might
warrant further study.


$\bf VI.\ Production\ Rate\ and\ Comparison\ to\ Galactic\ Sources$

The observed number of GRB is given by $N_{obs}\sim
10^{-7}{\rm {yr^{-1}galaxy^{-1}}}$.  
Because $\tau_{life}\gg {\rm 1\ yr}\gg\tau_{dur}$ for the precessing jet GRB, the required
production rate of the objects is much reduced from a 
model for which
$\tau_{life}\sim \tau_{dur}$ even though the emission is strongly beamed.
The reason is that 
an object would account for an event in a given year if 
the line of sight is within
the solid angle swept out during that year by the beam, 
instead of having to fall strictly within
a solid angle $\sim \gamma^{-2}$.
Since the repetition time is the time scale for which
the beam fills the full $4\pi$ solid angle, 
in 1 year, the beam would sweep through a solid angle 
$4\pi (1 {\rm yr}/\tau_{rep})$ in this time.  Thus the required production
rate is
$$N_{pro} \sim N_{obs}(\tau_{rep}/ 1{\rm yr})/4\pi \sim  10^{-5}
(\tau_{rep}/10^3{\rm yr}).\eqno(20)$$ 
In a model for which $\tau_{life}\sim \tau_{dur}$ then
$N_{pro} \sim \gamma^2 N_{obs}$ which is much larger and excessive.

 
Though the jets require blob forming instabilities and jet collimation
which are poorly understood phenomena, 
such features are definitively observed 
in active galactic nuclei (AGN) and micro-quasar jets GRS1915-105
and SS433 (Blandford, 1993; Vermeulen, 1993; Mirabel \& Rodriguez, 1994).  
In particular, the SS433 jets precess with $\tau_{pn1}\sim 10^7 {\rm sec}$,
with a secondary modulation of $\tau_{pn2}\sim 10^8 {\rm sec}$.
   Bullets and blobs of 
emission are seen distinctively, with $(13)$ satisfied, namely that
$\tau_{dec}\sim\om_{prod}^{-1}\sim\tau_{swp}\sim 10^5 {\rm sec}$.
Beamed emission with $\ga> 10^4$,  would be visible 
only for  $\tau_{swp}<10^3{\rm sec}$.  Any secondary PN of the jet
that moved it even $6\tm 10^{-4}$ radians would render it
invisible for a time scale longer than the lifetime of the rotational
decay of the source.  Note also that 
the SS433 blobs flow out from channels through either side of a thick
disk which may aid in the jet collimation \rl.


There would be about 100 jetted objects of this type
in the Galaxy at any one time (cf. $(7)$ and $(20)$),
which may not look significantly different from normal pulsars
from angles not pointed favorably to
see their gamma emission.
The gamma ray emitting  blobs in the GRB pulsars of the Galaxy
must have a minimum Lorentz factor 
such that the probability of the beam solid angle of passing
through our line of sight is  $<1/100$.
This gives an angle $\sim 1/(\pi\ga_{min}^{2})<1/100$ so that
$\ga_{min}>\sim 5.6$.  
Finally we note that a fraction of the objects, those associated
with SPR, would be
necessarily accompanied by a burst of gravitational radiation from the
stellar collapse and the unstable rotation.  The fraction of these objects
requires determination of what kind of PSR form from accretion induced
collapse (Yi \& Blackman, 1996).
Stellar mass micro-quasars may occur in
classes similar to AGN.  GRB may be a signature of one such class.

---------------------------------------

\noindent 1. Because of the short time scale of SPR, the physics
of the gamma-ray emission may be complicated by neutrino emission
(Thompson, 1994) 
in the context of the unipolar inductor approach used above, but
further discussion is outside of the scope of the present work. 



\noindent We thank R. Narayan for references, and 
C. Thompson and E.T. Vishniac  for comments.
 
\noindent Asseo, E., Kennel, C.F., 1978, \& Pellat,R., A. \& A., $\bf
65$, 401.

\noindent Blandford,R.D., 1993, in Burgarella et al., 1993.

\noindent Blackman, E.G., 1995, PhD. thesis, Department of Astronomy,
Harvard University.

\noindent Burgarella,D., Livo,M., \& O'Dea,C. eds., 1993, 
$Astrophysical\ Jets$, 
Space Telescope Symposium 6, (Cambridge: Cambridge Univ. Press).

\noindent Chandrasekhar, S., 1987 {\it Ellipsoidal Figures of Equilibrium},
(New York:  Dover).

\noindent Chiueh,T., Li, Z., \& Begelman M.C., 1991, Ap.J., $\bf 377$, 462.

\noindent Fargion, D. \& Salis, A., 1996, astro-ph SISSA preprint 9605166.

\noindent Fatuzzo,M.,\& Melia,F., 1993, Ap.J. $\bf 414$, L89. 

\noindent Fishman,G.J.,  Brainerd, J.J., \& Hurley,K eds., 1994,
$Gamma-Ray\ Bursts$, AIP Conference Proceedings 307. 

\noindent Friedman,J., 1983, Phys. Rev. Lett., $\bf 51$, 11.

\noindent Goldreich,P., \& Julian,W.H., 1969, Ap.J., $\bf 157$, 869.

\noindent Heyverts, J., \& Norman,C.A., 1989, Ap.J., $\bf 347$, 1055.


\noindent Kouveliotou,C., et al., 1994, in  Fishman et al., 1994.

\noindent Krolik,J.H. \& Pier,E.A. 1991, ApJ, 373, 277.


\noindent Meegan,C., et al., 1994, in Fishman et al., 1994.

\noindent Michel,F.C., Ap.J., 1984, $\bf 284$, 384.

\noindent Mirabel,I.F., \& Rodriguez,L.F., 1994, Nature, $\bf 371$, 46.

\noindent Narayan,R. \& Popham, R., Ap. J., 1989, $\bf 346$, L25.

\noindent Nemiroff,R.J., et al., 1994, in Fishman et al., 1994.

\noindent Ostriker,J.P. \& Gunn,J.E., 1969,  Ap.J., $\bf 157$, 1395.


\noindent Beskin, V.S., Gurevich, A.V. \& Istomin, Ya. N., 1993, 
{\it Physics of the Pulsar Magnetosphere}, (Cambridge: Cambridge University Press) p97.


\noindent Shapiro,S.L. \& Teukolsky,S.A., 1983, 
$Black\ Holes\ White\ Dwarfs\ \&\ Neutron\ Stars$, (New York: Wiley).

\noindent Thorne,K.S., Price,R.H., 1986, \& MacDonald,D.A., $Black\ Holes:\
The\ Membrane\ Paradigm,$ (New Haven: Yale Univ. Press).

\noindent Thompson, C., 1994, MNRAS, $270$, 480.

\noindent Usov,V.V., 1992, Nature, $\bf 357$, 452.

\noindent Usov,V.V., 1994, in  Fishman et al., 1994.

\noindent van den Heuvel,E.P.J., 1986, in $The \ Evolution\ of\
Galactic\ X-ray\ Binaries$, J. Truemper, W.H.G. Lewin, \&
W. Brinkmann eds., NATO ASI Series $167$, (Dodrecht: Reidel).

\noindent Vermeulen,R., 1993, in Burgarella et al., 1993.

\noindent Yi, I, \& Blackman, E.G., 1996, preprint.

\end